\documentclass{article}
\usepackage{bm,latexsym,amsmath,amssymb,amsfonts,fancyhdr,color,graphicx,multirow,cite}
\usepackage{subfigure}
\usepackage[a4paper,bottom=3cm,top=2.5cm,head=5mm,width=17cm,dvipdfm]{geometry}
\usepackage[usenames,dvipsnames,svgnames,table]{xcolor}
\usepackage[colorlinks=true,
          linkcolor=blue,
          urlcolor=blue,
           citecolor=green,          
]{hyperref}
\usepackage[dotinlabels]{titletoc}
\usepackage{hyperref}
\usepackage{titlesec}
\usepackage{authblk}
\usepackage{stackengine}
\graphicspath{{figs/}}

\begin{document}
\fontsize{12pt}{14pt}\selectfont

\title{
       \textbf{\fontsize{16pt}{18pt}\selectfont Broken Democracy with Intermediate $\mathbb S_2 \times \mathbb S_2$ Residual Symmetry and Random Perturbations}}
\author[1]{{\large Neil D. Barrie} \footnote{\href{mailto:neil.barrie@ipmu.jp}{neil.barrie@ipmu.jp}}}
\author[2]{{\large Shao-Feng Ge} \footnote{\href{mailto:gesf@sjtu.edu.cn}{gesf@sjtu.edu.cn}}}
\author[1,2]{{\large Tsutomu T. Yanagida} \footnote{\href{mailto:tsutomu.tyanagida@ipmu.jp}{tsutomu.tyanagida@ipmu.jp}}}
\affil[1]{\small Kavli IPMU (WPI), UTIAS, The University of Tokyo, Kashiwa, Chiba 277-8583, Japan}
\affil[2]{\small Tsung-Dao Lee Institute \& School of Physics and Astronomy, Shanghai
Jiao Tong University, Shanghai 200240, China}
\date{\today}

\maketitle

\vspace{-100mm}
\hfill IPMU19-0166
\vspace{95mm}

\begin{abstract}
\fontsize{11pt}{13pt}\selectfont
The democratic mass matrix is an intriguing possibility for explaining the observed
fermion mixings due to its inherent hierarchical mass eigenvalues and large mixing angles. 
Nevertheless, two of the three mass eigenvalues are zero if the flavor democracy
is exact, in obvious contradiction with the experimental observations.
One possibility is breaking the flavor democracy with anarchical perturbations as
we proposed in an earlier work. However, even within the first two generations,
the charged fermion masses are also hierarchical which may not be a coincidence.
The democratic $\mathbb S^L_3 \times \mathbb S^R_3$ symmetry of the three generations
may  first be  broken down to an intermediate $\mathbb S^L_2 \times \mathbb S^R_2$ symmetry
among the first two generations to regulate
the sequential hierarchies, followed by random perturbations, that generate the correct size of all measured observables. Unique predictions for neutrinoless double beta decay are also found.
\end{abstract}

\section{Introduction}
\label{sec:intro}

Both quarks and leptons have nontrivial mixing among the up- and down-type fermions, 
but their mixing patterns are very different. While the Cabibbo-Kobayashi-Maskawa
(CKM) matrix of quark mixing is quite close to the unit matrix, the
Pontecorvo-Maki-Nakagawa-Sakata (PMNS) matrix of neutrino mixing exhibits large
mixing angles. This difference might be indicative of some profound picture of the
fundamental theory. Theoretically speaking, quarks and leptons have the same charge
assignment under the $SU(2)_L$ weak group but different under the $SU(3)_c$
strong group and the $U(1)_Y$ hypercharge group. However, both $SU(3)_c$
and $U(1)_Y$ are gauge groups and as such have little to do with flavor mixing.
This might be true for $SU(3)_c$ but the interplay between $SU(2)_L$ and
$U(1)_Y$ leads to an important difference between quarks and leptons.
Namely, the lepton sector contains neutrally charged neutrinos while  all quarks
are charged, even after electroweak symmetry breaking. Consequently, neutrinos
can be Majorana type fermions and their mass terms can arise via the seesaw mechanism
in the presence of heavy right-handed Majorana neutrinos \cite{seesaw}. If all of the charged fermions exhibit similar mixing matrix structures
($V_u \sim V_u \sim V_\ell$, respectively), but neutrinos have a unique mixing structure
($V_\nu$), it would be natural to observe $V_{\textrm{CKM}} \equiv V^\dagger_u V_d \sim I$
while $V_{\textrm{PMNS}} \equiv V^\dagger_\ell V_\nu$ is quite different from the unit matrix
$I$. Thus, we may have already observed
evidence for the seesaw mechanism, in addition to grand unification theory (GUT). 
The flavour mixing paradigm considered here not only supports the seesaw mechanism, but also the existence of heavy right-handed Majorana neutrinos, the GUT, and neutrinoless double beta decay.

A fundamental understanding of the exact form of the individual mixing matrices ($V_u$,
$V_d$, $V_\nu$, and $V_\ell$) remains unknown.
The many existing attempts can be classified into two main categories:
the top-down and the bottom-up approaches
\cite{Weinberg:1977hb,Fritzsch:1977za,Fritzsch:1979zq,Tanimoto:1999pj,Fukugita:1998vn}.

\begin{enumerate}
\item In the top-down approach, initially a full flavor symmetry is imposed on
the fundamental Lagrangian before being broken,  providing predictions of the observed mixing angles.
Note that the full flavor symmetry has to be broken, otherwise the CKM (PMNS)
matrix becomes a trivial unit matrix since the left-handed up- and down-type
quarks (leptons) are in the same $SU(2)_L$ weak doublet and consequently are subject
to the same flavor symmetry and mass matrix structure. In this case, it is not obvious if
the full flavor symmetry really predicts the mixing pattern that arises after
symmetry breaking and the fermions obtaining mass. If the mixing pattern is truly
determined by symmetry, it has to be some residual symmetry that survives the
symmetry breaking.

\item The bottom-up approach starts from residual symmetries, either assumed
by ansatz or reconstructed from the measured mixing matrices
\cite{Lam:2008rs,Lam:2008sh}. The full flavor symmetry
is then reconstructed by taking a product group of the residual symmetries
\cite{Hernandez:2012ra,Hernandez:2012sk,Esmaili:2015pna}.

\end{enumerate}

In either scenario, the residual symmetries control the mixing patterns of the quark and lepton sectors
\cite{Dicus:2010yu,Ge:2011ih,Ge:2011qn,Fonseca:2014koa,Ge:2014mpa},
since the residual symmetries are the ones directly connected with the mixing matrix.
In particular, residual $\mathbb Z_2$ symmetries can predict unique
correlations among the mixing angles and the Dirac CP phase, which can
be readily tested by neutrino oscillation experiments. Similar correlations
also appear when considering perturbed by anarchy perturbations, in which the inherent $\mathbb S^L_3 \times \mathbb S^R_3$ symmetry of democracy regulates \cite{Ge:2018ofp} the otherwise completely random anarchy model  \cite{Hall:1999sn,Haba:2000be}. This improved
version of the anarchy model with assistance from residual symmetry takes advantage of the benefits
from both residual symmetries and random perturbations.

An intriguing feature of the democratic mass matrix
\cite{Tanimoto:1999pj,Fukugita:1998vn,Harari:1978yi,Koide:1989ds,Tanimoto:1989qh,
Kaus:1990ij,Fritzsch:1989qm,Fritzsch:1994yx,Branco:1995pw,Fritzsch:1995dj,
Xing:1996hi,Mondragon:1998gy,Fritzsch:1998xs,Fritzsch:1999ee,Haba:2000rf,
Branco:2001hn,Fujii:2002jw,Fritzsch:2004xc,Rodejohann:2004qh,Teshima:2005bk,
Xing:2010iu,Zhou:2011nu,Dev:2012ns,Canales:2012dr,Jora:2012nw,Yang:2016esx,
Fritzsch:2017tyf,Si:2017pdo} is its inherent mass hierarchy and large values of
the $\theta_{12}$ and $\theta_{23}$ mixing angles. The corresponding democratic
$ \mathbb S^L_3 \times \mathbb S^R_3 $ symmetry dictates only one non-zero eigenvalue,
$m_1 = m_2 = 0$ while $m_3 \neq 0$,
which nicely explains why the third-generation fermions have the largest mass.
In this context, the smallness of the masses of the other fermions, $m_{1,2} \ll m_3$,
may be due to them originating from perturbative corrections.
Starting with democratic mass matrices for the quarks and charged leptons, and
a diagonal Majorana neutrino mass matrix, we have $V_{\textrm{CKM}}=I$ and
$ V_{\textrm{PMNS}} = V^\dagger_\ell$. Upon introducing perturbative corrections, the observed fermion masses and mixing angles can then be explained
\cite{Ge:2018ofp,Ghosh:2018tzv}. As global symmetries are not protected against
the quantum gravity effects, the perturbations are naturally anarchical.
\\

Although the $\mathbb S^L_3 \times \mathbb S^R_3$ model
naturally explains $m_{1,2} \ll m_3$, there is no prediction of $m_1 \ll m_2$.
The sequential hierarchy among fermion masses, $m_1 \ll m_2 \ll m_3$
for the three generations of charged fermions, might be an indication of sequential breaking
of residual symmetries. We will show that by first breaking
$\mathbb S^L_3 \times \mathbb S^R_3$ down to $\mathbb S^L_2 \times \mathbb S^R_2$ \cite{Fritzsch:1989qm,Ghosh:2018tzv}
and then  adding small random perturbations to break all global flavour symmetries, the observed sequential hierarchies can be naturally explained.
In other words, the residual symmetries can further divide
into different groups or separate at sequential steps. Some attempts have been made to consider different residual symmetries in both the
lepton and quark sectors \cite{Ge:2014mpa} with each sector containing unique divisions of residual symmetries. In this paper we provide a sequential division of residual
symmetries common to the charged fermions.

In Section \ref{sec:broken}, we first show analytically the implications of breaking
the democratic residual symmetry $\mathbb S^L_3 \times \mathbb S^R_3$ down to
a reduced version of the $\mathbb S^L_2 \times \mathbb S^R_2$
residual symmetry. Then in Section \ref{sec:random}, we introduce random perturbations
to break the remaining $\mathbb S^L_2 \times \mathbb S^R_2$ residual symmetries
to allow nonzero mass eigenvalues for all fermions and at the same time
obtain realistic values of the CKM and PMNS mixing angles. Section \ref{sec:nuDBD} discusses the predictions for neutrinoless double beta decay within our model. Our conclusions are
summarized in Section \ref{sec:conclusion}.

\section{Broken Democracy and Intermediate $ \mathbb S^L_2 \times \mathbb S^R_2 $ Residual Symmetry}
\label{sec:broken}

The democratic mass matrix, dictated by the $\mathbb S^L_3 \times \mathbb S^R_3$
symmetry, can provide a natural explanation for the large hierarchy between the
third generation fermions and their counterparts in the first two generations.
The flavor democracy is realized within the Standard Model through explicit
breaking of the globally continuous $O(3)_L\times O(3)_R$ symmetry by the democratic
vacuum expectation value (VEV) pattern of the triplet scalars
$\langle\phi_{L,R} \rangle=(1,1,1)$,
with the corresponding Yukawa interaction terms,
\begin{equation}
	y (\bar{\psi}_{L,i} \phi_{L,i}) (\phi_{R,j} \psi_{R,j}) ~,
\label{yuk1}
\end{equation}
for all the charged fermions. The resulting rank-3 democratic mass matrix is generally \\
parametrised as,
\begin{equation}
  M_f^{(0)}
=
  \frac{M_{3f} }{3}
\begin{pmatrix}
1 & 1 & 1   \\
1 & 1 & 1   \\
1 & 1 & 1  
\end{pmatrix} \,,
\label{eq:Mf0}
\end{equation}
with $M_{3f}$ denoting the only non-zero eigenvalue $M_f^{(0)}$. Since an overall
phase can be easily rotated away and has no physical consequences, we take all matrix
elements of Eq. \ref{eq:Mf0} to be real.
The diagonalization of Eq. \ref{eq:Mf0}, $M_f = V_L D_f V^\dagger_R$,
involves two mixing matrices, $V_L$ and $V_R$ for the left- and right-handed
fermions, respectively;  where $V_R^{\dagger} = V_L $ as the mass matrix is symmetric. In general, since the physical CKM and PMNS mixing matrices are
only for left-handed fermions, diagonalizing $M_f M^\dagger_f = V_L D_f^2 V^\dagger_L$
instead can drop out $V_R$, leaving only $V_L$. For the democratic mass matrix, Eq. \ref{eq:Mf0},
one possibility of the mixing matrix $V_L$ is,
\begin{equation}
	V^\dagger_L
=
\begin{pmatrix}
  1 \\
& \frac 1 {\sqrt 3} & \frac {-2}{\sqrt 6} \\
& \frac 2 {\sqrt 6} & \frac 1 {\sqrt 3}
\end{pmatrix}
\begin{pmatrix}
  \frac 1 {\sqrt 2} & \frac {-1} {\sqrt 2} \\
  \frac 1 {\sqrt 2} & \frac 1 {\sqrt 2} \\
& & 1
\end{pmatrix}
=
\begin{pmatrix}
	\frac 1 {\sqrt 2} & \frac {-1} {\sqrt 2}    & 0 \\
	\frac 1 {\sqrt 6} & \frac 1 {\sqrt 6} & \frac {-2} {\sqrt 6} \\ 
  \frac 1 {\sqrt 3} & \frac 1 {\sqrt 3} & \frac 1 {\sqrt 3}
\end{pmatrix} \,.
\label{eq:UL}
\end{equation}
In the standard parametrisation of the CKM and PMNS matrices,
\begin{equation}
  \mathcal P
\begin{pmatrix}
  1 \\
& c_{23} & s_{23} \\
&-s_{23} & c_{23}  
\end{pmatrix}
\begin{pmatrix}
  c_{13} & & s_{13} e^{- i \delta_D} \\
& 1 & \\
- s_{13} e^{i \delta_D} & & c_{13}
\end{pmatrix}
\begin{pmatrix}
  c_{12} & s_{12} \\
- s_{12} & c_{12} \\
  & & 1
\end{pmatrix}
  \mathcal Q \,,
\label{eq:V-parametrization}
\end{equation}
where $\delta_D$ is the Dirac CP phase while
$\mathcal P \equiv \mbox{diag}\{e^{i \alpha_1}, e^{i \alpha_2}, e^{i \alpha_3}\}$
and
$\mathcal Q \equiv \mbox{diag}\{e^{i \phi_1}, e^{i \phi_2}, e^{i \phi_3}\}$
are rephasing matrices, the mixing angles take the values,
\begin{equation}
	\theta^f_{13}
=
  0^{\circ},
\qquad
	\theta^f_{12}
=
  45^{\circ},
\quad \mbox{and} \quad
  \theta^f_{23}
=
  54.7^{\circ} \,,
\label{dem_angles}
\end{equation}
together with $\mathcal P = \mathcal Q = \{1, -1, 1\}$.
Thus, the democratic flavor structure induces large $\theta^f_{12}$ and $\theta^f_{23}$ mixing angles
as observed in the PMNS matrix. Correspondingly, the three mass eigenvalues are
$m_1 = m_2 = 0$ and $m_3 = M_{3f}$. Note that the mixing matrix in Eq.  \ref{eq:UL} is
not unique due to the two-fold degeneracy between $m_1$ and $m_2$. A $2 \times 2$
mixing among the degenerate eigenstates modifies all the mixing angles and the
Dirac CP phase, but leaves a unique correlation among them, as elaborated in 
Section 2 of \cite{Ge:2018ofp}. The mixing pattern in Eq. \ref{eq:UL} applies to all
charged fermions, naturally leading to zero 1-3 and 2-3 mixings in the quark
sector.

To obtain non-zero 1-3 and 2-3 quark mixings as well as non-zero fermion masses for
the first two generations, additional contributions are necessary. Since the
masses of the first two generations are also hierarchical, they might have different
origins. In other words, it is more appropriate to first allow one of them to be
non-zero while keeping the other one zero. A natural realization of this is breaking the
democratic $\mathbb S^L_3 \times \mathbb S^R_3$ symmetry to a smaller democratic
$\mathbb S^L_2 \times \mathbb S^R_2$ symmetry under which the mass matrix in Eq. \ref{eq:Mf0}
receives the following correction term,
\begin{equation}
	M_f^{(1)}
=
	M_{3f} \times b_f e^{i\beta_f}
\begin{pmatrix}
	1 & 1 & 0   \\
	1 & 1 & 0   \\
	0 & 0 & 0 
\end{pmatrix} \,,
\label{mass1}
\end{equation}
where $\beta_f$ is an overall random phase. As this is a perturbation to the leading order
democratic mass matrix in Eq. \ref{eq:Mf0}, the real parameter $b_f$ should be smaller
than $ \frac{1}{3} $. This rank-2 democratic perturbation can naturally arise from a yukawa interaction term containing the two triplet scalars
$\phi^{\prime}_{L,R}$,
\begin{equation}
  y^{\prime}
  (\bar{\psi}_{L,i} \phi^{\prime}_{L,i})
  (\phi^{\prime}_{R,j} \psi_{R,j}) \,,
\label{yuk2}
\end{equation}
in the same way as in Eq. \ref{yuk1}
via a rank-2 democratic VEV $\langle\phi^{\prime}_{L,R} \rangle=(1,1,0)$.

After diagonalization, two non-zero masses appear with the ratio,
\begin{equation}
  \frac{m_2}{m_3}
=
  \frac {2 b_f}
        {3 + 8 b_f \cos\beta_f + 12 b_f^2}
\approx
  \frac 2 3 b_f \,,
\label{mass2}
\end{equation}
where $m_2$ and $m_3$ are the non-zero eigenvalues of the mass matrix
$M_f^{(0)}+ M_f^{(1)}$. The hierarchy among $m_2$ and $m_3$ is almost
uniquely controlled by $b_f$ and slightly by $\beta_f$ if $b_f$ is not
very small.

The mixing angles can also be derived analytically,
\begin{equation}
  \sin^2 \theta^f_{13}
=
  0,
\quad
  \sin^2 \theta^f_{12}
=
  \frac 1 2,
\quad
  \sin^2 \theta^f_{23}
\approx
  \frac 2 3
\left(
  1
+ \frac 4 3 b_f \cos \beta_f
\right) \,.
\label{angle_prediction}
\end{equation}
When expanded to the linear order, the mixing angles are, $\theta^f_{13} = 0$,
$\theta^f_{12} = 45^\circ$, and $\theta^f_{23} \approx 54.7^\circ +
\frac 2 3 \sqrt 2 b_f \cos \beta_f$ together with
$\mathcal P = \mathcal Q = \{1,-1,1\}$. In addition to real mixing angles,
the diagonalization matrix also contains a complex rephasing term,
\begin{equation}
  V^\dagger_f
=
\begin{pmatrix}
  1 \\
& \cos \theta^f_{23} & \sin \theta^f_{23} \\
&-\sin \theta^f_{23} & \cos \theta^f_{23}
\end{pmatrix}
\begin{pmatrix}
  1 \\
& 1 \\
& & e^{i \phi_f}
\end{pmatrix}
\begin{pmatrix}
  \cos \theta^f_{12} & \sin \theta^f_{12} \\
- \sin \theta^f_{12} & \cos \theta^f_{12} \\
  & & 1
\end{pmatrix} \,,
\label{eq:V}
\end{equation}
where the phase is suppressed by $ b_f $, $\phi_f \approx 2 b_f \sin \beta_f$.
It is interesting to see that the addition of the $\mathbb S^L_2 \times \mathbb S^R_2$ term only  perturbs the 2-3
mixing while keeping the 1-2 and 1-3 mixings unaffected. This leads to very unique features compared to those explored in our previous paper \cite{Ge:2018ofp}.

\subsection{Quark Sector}

Using the derived analytical equations derived above, in combination with the measured quark masses, we can estimate the typical
sizes of the $b_f$ parameters. For up-type quarks, 
$b_u \approx 3 m_c / 2 m_t = 0.0056$
(with $m_c(M_Z) = 0.638~\mbox{GeV}$ and $m_t(M_Z) = 172.1\,\mbox{GeV}$) while
for down-type quarks,
$b_d \approx 3 m_s / 2 m_b = 0.03$
(with $m_s(M_Z) = 57~\mbox{MeV}$ and $m_b(M_Z) = 2.86\,\mbox{GeV}$); each considering the central values at the $M_Z$ scale \cite{Xing:2011aa}.

The mixing pattern in Eq. \ref{eq:V} applies for both up- and down-type quarks.
Combining the up and down quark mixings, $V_u$ and $V_d$, the predicted CKM matrix
is dominated by 2-3 mixing,
\begin{equation}
  V_{\textrm{CKM}}
\equiv
  V^\dagger_u V_d
=
\begin{pmatrix}
  1 & 0 & 0 \\
  0 & \cos \theta^{\textrm{CKM}}_{23} & \sin \theta^{\textrm{CKM}}_{23} \\
  0 &-\sin \theta^{\textrm{CKM}}_{23} & \cos \theta^{\textrm{CKM}}_{23}
\end{pmatrix} \,,
\end{equation}
where
\begin{equation}
  \left| \theta^{\textrm{CKM}}_{23} \right|
\approx
  \frac {2 \sqrt 2} 3
  \sqrt{b^2_u - 2 b_u b_d \cos(\beta_u - \beta_d) + b^2_d} \,~.
\end{equation}
The 2-3 mixing angle varies in the range
$\left| \theta^{\textrm{CKM}}_{23} \right| \lesssim |b_u + b_d| \approx 2.04^\circ$, with this dependent upon the randomised $\beta_u$ and $\beta_d$ and the mass values of $m_s$ and $m_c$ considered, with the large uncertainty in the strange quark mass allowing greater maximum values of $\theta^{\textrm{CKM}}_{23}$.
Once considering the  constraints from  the measured fermion mass ratio, $\theta^{\textrm{CKM}}_{23}$ should be
of the percentage level which is exactly what we need to explain the experimental data,
$\theta^{\textrm{CKM}}_{23} = 2.38^{\circ +0.18}_{~-0.17} $ .
The 1-2 and 1-3 mixing angles will arise from the random perturbations to be discussed in Section \ref{sec:random}.

\subsection{Neutrino Sector}

While the 2-3 mixing is suppressed in the CKM matrix, it is close to the
maximal value in the PMNS matrix.
From the muon and tau masses, $m_\mu = 105.7~\mbox{MeV}$ and
$m_\tau = 1776.86\,\mbox{MeV}$, we can estimate the value of
$b_\ell \approx 3 m_\mu / 2 m_\tau = 0.089$, which corresponds to 
the atmospheric angle being $\theta_a \equiv \theta_{23} \sim 50.6 ^{\circ}$ for $ \beta_f=\frac{\pi}{2}$; given the size of $b_l$ the dependence on $\beta_l$ will have an important contribution.
Interestingly, in both the quark and lepton sectors, simply
requiring consistency with the observed mass ratios of the second and
third generation quarks and charged leptons, with an
$\mathbb S^L_2 \times \mathbb S^R_2$ symmetry, leads to predictions
for the CKM and PMNS $\theta_{23}$ mixing angles within the current
experimental $3 \sigma$ ranges. Although, the solar angle
$\theta_s \equiv \theta_{12} = 45^\circ$ takes the maximal value
while the reactor angle $\theta_r \equiv \theta_{13} = 0^\circ$ vanishes,
both deviating from their measured values by roughly $10^\circ$ within the lepton sector.

Since the charged lepton mixing already follows the pattern in Eq. \ref{eq:V}, at
the leading order, the only thing we can manipulate is the neutrino counterpart.
As argued in Section \ref{sec:intro}, the neutrino sector may not be subject to
the same mass matrix pattern as the charged fermion sectors because
neutrinos can be Majorana type fermions and can obtain their masses from the seesaw
mechanism. In the presence of only left-handed neutrinos, the neutrino mass matrix
does not follow the democratic $\mathbb S^L_3 \times \mathbb S^R_3$ symmetry but
a more restrictive $\mathbb S^L_3$ ($\times \mathbb S^L_3$) symmetry from both
sides. Then, we have two $ \mathbb S_3^L $ invariants,
\begin{equation}
M_\nu=
\begin{pmatrix}
	a & 0 & 0   \\
	0 & a & 0   \\
	0 & 0 & a
\end{pmatrix}
\quad
~~\textrm{or}~~
\quad
\begin{pmatrix}
0 & b & b   \\
b & 0 & b   \\
b & b & 0
\end{pmatrix} \,.
\label{eq:Mnu}
\end{equation}
While the first term of Eq. \ref{eq:Mnu} has no effect on the PMNS mixing, the second is
actually equivalent to Eq. \ref{eq:Mf0} for determining the mixing pattern.
To obtain a different mixing pattern than that of the charged fermions, no appearance of
the second form of Eq. \ref{eq:Mnu} can be allowed. In addition, the latest
global fit and cosmological constraints point to the neutrinos exhibiting a normal hierarchy (NH)
and small mass scales. The first of Eq. \ref{eq:Mnu} unfortunately dictates
a degenerate mass hierarchy. So both forms in Eq. \ref{eq:Mnu} should be
forbidden. This provides
another reason why the $\mathbb S^L_3 \times \mathbb S^R_3$ symmetries must be broken, in addition to the fermion mass hierarchy argument.  To alleviate the issues caused by these two terms, we impose a discrete $\mathbb Z_4$ symmetry under which the left handed and right handed neutrinos have charge $ +1 $ and $ -1 $ respectively, while $ \phi $ and $ \phi^{\prime} $ are neutral. This forbids neutrino mass terms of the form present in Eq. (\ref{eq:Mnu}) to be generated by the scalars $\phi $ and $ \phi^{\prime} $. 

Regardless of the ultraviolet (UV) completion, neutrinos receive
their mass term from the Weinberg-Yanagida operator \cite{Weinberg,Yanagida},
\begin{equation}
  y^{\prime\prime}
\left( \psi_{L,i} \phi^{\prime\prime}_i \right)
\left( \phi^{\prime\prime}_j \psi_{L,j} \right) HH \,,
\label{yuk3}
\end{equation}
where we have introduced the scalar $ \phi^{\prime\prime}_i $ which carries $\mathbb Z_4$ charge $ +1 $, such that the term is $\mathbb Z_4$ symmetric. To achieve this, the  fields $\psi_L$
and $\phi''$ transform as $\psi_L \rightarrow i \psi_L$ and
$\phi'' \rightarrow i \phi''$ under the imposed $\mathbb Z_4$ symmetry, while all other fields are invariant.
When $\phi''$ develops the VEV structure  $\langle \phi'' \rangle \propto (0,0,1)$,
the Majorana neutrino mass matrix becomes,
\begin{equation}
M_\nu^{(0)}=
M_{3\nu}\begin{pmatrix}
0 & 0 & 0   \\
0 & 0 & 0   \\
0 & 0 & 1
\end{pmatrix} \,,
\label{neutrino_mass1}
\end{equation}
instead of Eq. \ref{eq:Mnu}. This only contains one non-zero mass eigenvalue,
with anarchy contributions to generate the remaining masses, naturally leading
to a normal hierarchy.

The degeneracy of the two vanishing mass eigenvalues in Eq. \ref{neutrino_mass1}
leads to arbitrary 1-2 rotations. Consequently, the PMNS matrix becomes,
\begin{equation}
\hspace{-3mm}
  V_{\textrm{PMNS}}
=
\begin{pmatrix}
  1 \\
& c^\ell_{23} & s^\ell_{23} \\
&-s^\ell_{23} & c^\ell_{23}
\end{pmatrix}
\hspace{-1mm}
\begin{pmatrix}
  c^\ell_{12} & s^\ell_{12} \\
- s^\ell_{12} & c^\ell_{12} \\
  & & 1
\end{pmatrix}
\hspace{-1mm}
\begin{pmatrix}
  c_T & - s_T e^{i \phi} \\
  s_T e^{- i \phi} & c_T \\
& & 1
\end{pmatrix}
\hspace{-1mm}
\begin{pmatrix}
  e^{i \alpha_1} \\
& e^{i \alpha_2} \\
& & e^{i \bar \alpha_3}
\end{pmatrix},
\end{equation}
where the $\phi_\ell$ in Eq. \ref{eq:V} has been combined into
$\bar \alpha_3 \equiv \alpha_3 + \phi_\ell$. In the presence of these extra degrees
of freedom, the solar angle can easily adjust while the atmospheric
and reactor angles remain unaffected. The value of the extra mixing
angle $\theta_T$ and CP phase $\phi$ are subject to the random
perturbations that we will elaborate on in Section \ref{sec:random}.

\section{Random Perturbations}
\label{sec:random}

After breaking democracy with $\mathbb S^L_2 \times \mathbb S^R_2$ perturbations,  the first generation charged fermions still remain massless and hence extra contributions are required. To do this we introduce additional small anarchy perturbations to each of the fermionic mass matrices. Given the mass hierarchies within the fermion sector, for these perturbations to explain the light fermion masses they must be small relative to $ M^{(0)} $ and $ M^{(1)} $. The anarchy perturbations we introduce are of the following form,
\begin{equation}
	M_f^{(2)}=
	M_{3f}
\begin{pmatrix}
	{\tilde \epsilon}_{11}^{f} & {\tilde \epsilon}_{12}^{f} & {\tilde \epsilon}_{13}^{f}   \\
	{\tilde \epsilon}_{21}^{f} & {\tilde \epsilon}_{22}^{f} & {\tilde \epsilon}_{23}^{f}   \\
	{\tilde \epsilon}_{31}^{f} & {\tilde \epsilon}_{32}^{f} & {\tilde \epsilon}_{33}^{f}  
\end{pmatrix}
\end{equation}
where $ {\tilde \epsilon}^f_{ij}= \epsilon^f_{ij}e^{i \alpha^f_{ij}} $ with the random values taken within the ranges $ \epsilon^f_{ij}\in[0,\epsilon^f_{\textrm{max}}] $ and $ \alpha^f_{ij}\in[0,2\pi] $. This contribution violates the global flavour symmetries within the model, and may have its origin in quantum gravity effects. Combining this with $ M^{(0)} $ and $ M^{(1)} $ we arrive at the full mass matrix for the charged fermions,
\begin{equation}
\frac{M_f}{M_{3f}}=
\frac{1}{3}
\underbrace{\begin{pmatrix}
1 & 1 & 1   \\
1 & 1 & 1   \\
1 & 1 & 1  
\end{pmatrix}}_{\mathbb S^L_3 \times \mathbb S^R_3}
\quad
+
\quad
e^{i\beta_f}
\underbrace{\begin{pmatrix}
	b_f & b_f & 0   \\
	b_f & b_f & 0   \\
	0 & 0 & 0 
\end{pmatrix}}_{\mathbb S^L_2 \times \mathbb S^R_2}
\quad
+
\quad
	\underbrace{\begin{pmatrix}
{\tilde \epsilon}_{11}^{f} & {\tilde \epsilon}_{12}^{f} & {\tilde \epsilon}_{13}^{f}   \\
	{\tilde \epsilon}_{21}^{f} & {\tilde \epsilon}_{22}^{f} & {\tilde \epsilon}_{23}^{f}   \\
	{\tilde \epsilon}_{31}^{f} & {\tilde \epsilon}_{32}^{f} & {\tilde \epsilon}_{33}^{f}  
\end{pmatrix}}_{\textrm{Anarchy}} 
\label{mass}
\end{equation}
where we have indicated the symmetry properties of each contribution. From here we consider the mixing angle predictions in both the lepton and quark sectors when requiring consistency of mass predictions with experimental observation. The analytical result in Eq. (\ref{mass2}) shall be utilised to obtain a relation between $b_f$ and $\beta_f$ to fix the correct second generation fermion masses, namely,
\begin{equation}
  b_f=\frac{1}{12 r_f^0}  (1-4 r_f^0 \cos(\beta_f))\left(1 -\sqrt{1-\left(\frac{  6 r_f^0}{1-4 r_f^0 \cos(\beta_f)}\right)^2} \right) 
  \label{rel2}
\end{equation}
where $r_f^0$ is the initial mass ratio of the second and first generation fermions prior to the introduction of anarchy perturbations.

\subsection{Lepton Sector}

The  PMNS matrix is constructed from the charged lepton and neutrino mixing matrices as follows, $ V_{\textrm{PMNS}}=V^{\dagger}_\ell V_\nu $. Given the observation of large mixing angles within the PMNS matrix the two leptonic mixing matrices must exhibit different mixing patterns, and hence forms of their mass matrices; unlike in the quark sector. In the following analysis we shall identify the charged lepton matrix $ M_l $ with that given in Eq. (\ref{mass}), and consider Majorana type neutrinos.  The initial neutrino mass matrix does not contain the democratic pattern, and is instead given by Eq. (\ref{neutrino_mass1}), due to its inherent $ \mathbb{Z}_4 $ symmetry. Without breaking the $ \mathbb S^L_2\times \mathbb S^L_2 $ symmetry in the neutrino mass matrix, the neutrino mass difference $ \Delta m_{21}^2 $ remains zero, and thus anarchy perturbations are introduced in an analogous way to the charged fermions.  The neutrino mass matrix is given by,
\begin{equation}
\frac{M_\nu}{M_{3\nu}}=
\underbrace{\begin{pmatrix}
0 & 0 & 0   \\
0 & 0 & 0   \\
0 & 0 & 1
\end{pmatrix}}_{\mathbb S^L_2\times \mathbb S^L_2 }
\quad
+
\quad
\underbrace{\begin{pmatrix}
{\tilde \epsilon}_{11}^{\nu} & {\tilde \epsilon}_{12}^{\nu} & {\tilde \epsilon}_{13}^{\nu}   \\
{\tilde \epsilon}_{12}^{\nu} & {\tilde \epsilon}_{22}^{\nu} & {\tilde \epsilon}_{23}^{\nu}   \\
{\tilde \epsilon}_{13}^{\nu} & {\tilde \epsilon}_{23}^{\nu} & {\tilde \epsilon}_{33}^{\nu}  
\end{pmatrix}}_{\textrm{Anarchy}}
\label{neutrino_mass}
\end{equation}
where the anarchy perturbations introduced here are symmetric due to the  Majorana nature of the neutrinos. The anarchy perturbations violate all the global symmetries of the model,  including the $\mathbb Z_4$ symmetry. Fixing the maximum of the range of the anarchy perturbations $\epsilon^{\nu}_{\textrm{max}}$ determines the ratio of the two mass differences $ \Delta m_{21}^2 $ and $ \Delta m_{31}^2 $ and hence can be constrained by observation. The value of $\epsilon^{\nu}_{\textrm{max}}$ also plays a key role in the predicted mixing angles of the PMNS matrix, as the anarchy perturbations are the sole contribution to $\theta_r$ and lead to smearing of $\theta_a$ and $\theta_s$ around the central value predicted by the $ \mathbb S^L_2\times \mathbb S^R_2 $ symmetry in the charged lepton sector. It should be noted that the maximum magnitude of the anarchy terms introduced to the neutrino sector appear relatively large compared with those for the charged fermions. This is to be expected, due to the suppression of the equivalent leading $ \mathbb S_3 $ symmetry term being disallowed by the $ \mathbb Z_4 $ symmetry.

\begin{figure}
	\centering
	\begin{subfigure}
		\centering
			\topinset{\bfseries(a)}{\includegraphics[width=0.45\textwidth]{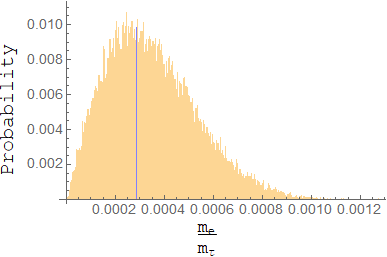}}{0.05in}{0.2in}
	\end{subfigure}
	\hfill
	\begin{subfigure}
		\centering
			\topinset{\bfseries(b)}{\includegraphics[width=0.45\textwidth]{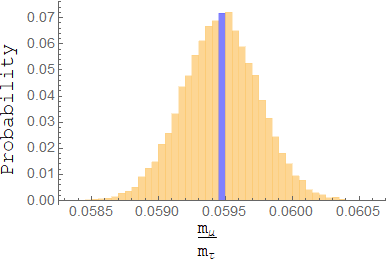}}{0.05in}{-.8in}
	\end{subfigure}
	\begin{subfigure}
	\centering
		\topinset{\bfseries(c)}{\includegraphics[width=0.45\textwidth]{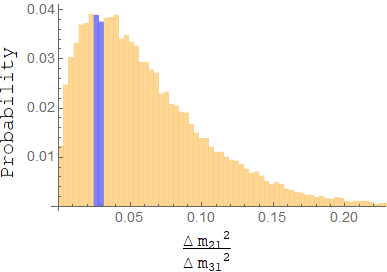}}{0.05in}{.2in}
	\end{subfigure}
	\caption{Masses of the (a) electron and (b) muon in units of the tau mass, and (c) ratio of the neutrino mass differences.  The blue regions indicate the mass ranges consistent with experiment within $3\sigma$ \cite{Tanabashi:2018oca}. The input parameters for this case are $  r^0_l=0.05925 $, $ \epsilon^l_{\textrm{max}}=0.0007 $, and $ \epsilon^\nu_{\textrm{max}}=0.28 $, with the angle $ \beta_l $ randomly selected within the range $ [0,2\pi] $.}  
\label{mass_l}
\end{figure}

Given that the masses of the charged leptons are well measured, the input parameters of the mass matrix are strongly constrained. The muon mass is negligibly effected by the anarchy perturbations,  due to the smallness of the electron mass, but this is sufficient to shift the mass out of the current $3\sigma$ range. Despite this it should be noted that the mixing angles are effectively independent of the electron mass, due to the small magnitude of the anarchy perturbations relative to the $ \mathbb S^L_2 \times \mathbb S^R_2 $ terms. 

\begin{figure}
	\centering
	\begin{subfigure}
		\centering
		\topinset{\bfseries(a)}{\includegraphics[width=0.45\textwidth]{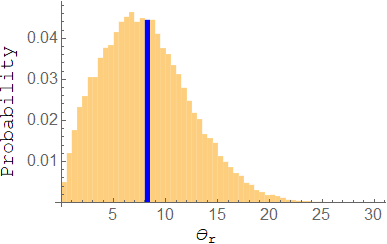}}{0.05in}{-.9in}
	\end{subfigure}
	\hfill
	\begin{subfigure}
		\centering
		\topinset{\bfseries(b)}{\includegraphics[width=0.45\textwidth]{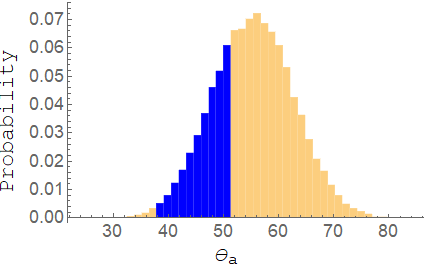}}{0.05in}{-.9in}
	\end{subfigure}
	\begin{subfigure}
		\centering
		\topinset{\bfseries(c)}{\includegraphics[width=0.45\textwidth]{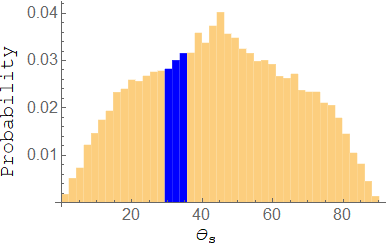}}{0.05in}{-.9in}
	\end{subfigure}
	\hfill
	\begin{subfigure}
		\centering
		\topinset{\bfseries(d)}{\includegraphics[width=0.45\textwidth]{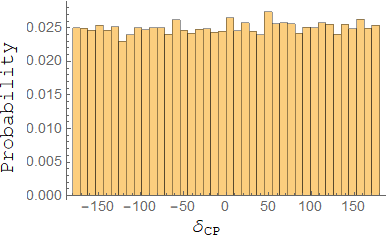}}{0.0in}{-.83in}
	\end{subfigure}
	\caption{Predicted mixing angles $\theta_{r}$, $\theta_{a}$, $\theta_{s}$, and $\delta_\mathcal{CP}$ phase in the lepton sector, for the mass distributions given in Fig. \ref{mass_l}, where $  (\theta_{a}, \theta_{r}, \theta_{s}) = (\theta_{23}, \theta_{13}, \theta_{12}) $.  The blue regions indicate the angle ranges consistent with experiment within $3\sigma$ \cite{Tanabashi:2018oca}. Note, the 3$ \sigma $ constraints on the $\delta_\mathcal{CP}$ phase have not been included.}
	\label{angles_l}
\end{figure}

We shall fix the size of the $ \mathbb S^L_2 \times \mathbb S^R_2 $ and anarchy perturbations such that the predicted ranges of the charged lepton and neutrino masses are consistent with experiment. This is depicted in Fig. \ref{mass_l}. The charged lepton anarchy perturbations determine the electron mass distribution, with the highest likelihood  mass approximately consistent with the observed $ m_e $ for $ \epsilon^l_{\textrm{max}}=0.0007$. In the neutrino sector, the anarchy perturbations determine the mass differences $ \Delta m_{21}^2$ and $\Delta m_{31}^2 $. In order to test consistency with observation we consider the ratio of the two measured mass differences, $ \frac{\Delta m_{21}^2}{\Delta m_{31}^2} $, and find compatibility between observation and the highest likelihood mass difference ratio when considering  $\epsilon^{\nu}_{\textrm{max}}=0.28 $. Now that the mass matrices are fully determined we can calculate the associated predictions of the PMNS mixing angles. The mixing angle predictions are presented in Fig. \ref{angles_l}, with current $3 \sigma$ constraints.

The PMNS mixing angle predictions obtained are found to be able to accommodate current experimental constraints. The most experimentally constrained of the three mixing angles $ \theta_{r} $ angle is sourced only by the neutrino anarchy perturbations. Interestingly, within this framework the predicted $ \theta_{r} $ peaks near the observed value when fixing the neutrino mass ratio, and hence maximum range of the anarchy perturbations  $\epsilon^f_{\textrm{max}}$, to the experimental value.  The effect of the anarchy perturbations on the $ \theta_a $ and $ \theta_{s} $ predictions is to produce broad distributions centred around the mixing angles predicted in Eq. (\ref{angle_prediction}), which are able to accommodate the observed mixing parameters. No preference for $ \delta_{\mathcal{CP}} $ is observed in either case.

\subsection{Quark Sector}

\begin{figure}
\centering
	\begin{subfigure}
		\centering
		\topinset{\bfseries(a)}{\includegraphics[width=0.45\textwidth]{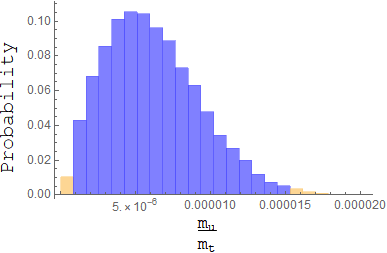}}{0.05in}{-.895in}
	\end{subfigure}
		\hfill
	\begin{subfigure}
		\centering
	\topinset{\bfseries(b)}{\includegraphics[width=0.45\textwidth]{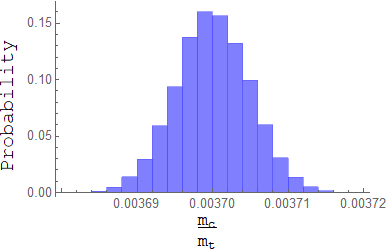}}{0.06in}{-.89in}
	\end{subfigure}
\centering
	\begin{subfigure}
	\centering
		\topinset{\bfseries(c)}{\includegraphics[width=0.45\textwidth]{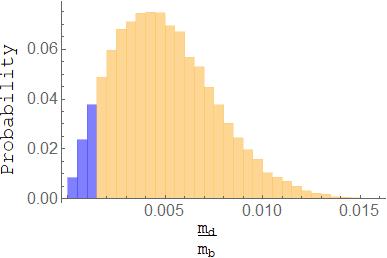}}{0.06in}{-.865in}
	\end{subfigure}
		\hfill
	\begin{subfigure}	
	\centering
		\topinset{\bfseries(d)}{\includegraphics[width=0.45\textwidth]{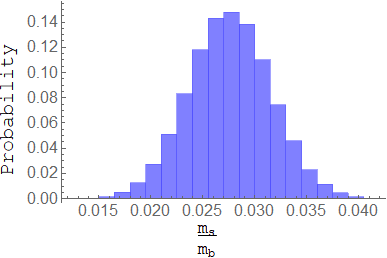}}{0.06in}{-.865in}
	\end{subfigure}
	
	\caption{Masses of the (a) up and (b) charm quarks in units of the top quark mass, and of the  (c) down and (d) strange quarks in units of the bottom quark mass, considered at the $M_Z$ scale \cite{Xing:2011aa}.  The blue regions indicate the mass ranges consistent with experiment within $3\sigma$ \cite{Tanabashi:2018oca}. The input parameters for this case are $   r_c^0=0.0037 $, $\epsilon^u_{\textrm{max}}=0.000012 $, $  r_d^0=0.026 $,   and $ \epsilon^d_{\textrm{max}}=0.01 $, with  $\beta_u$ and  $\beta_d$ individually randomised in the range $[0, 2\pi]$.  }
	\label{mass_q}
\end{figure}

We shall now investigate the model predictions for the CKM matrix, which is constructed from the quark mixing matrices through $ V_{\textrm{CKM}} = V_u^{\dagger}V_d $. In what follows, the up and down quark sectors will both have mass matrix patterns of the form given in Eq. (\ref{mass}). The $ \mathbb S^L_2 \times \mathbb S^R_2 $ contributions to the mass matrices determine the masses of the charm and strange quarks, and will be fixed by the relation given in Eq. (\ref{rel2}). The width of the distributions of the second generation quark masses is determined by the size of the anarchy perturbations. The anarchy perturbations also introduce the first generation quark masses and provide the sole contributions to the CKM mixing angles  $\theta_{13}$ and $\theta_{12}$. The predicted quark masses are presented in Fig. \ref{mass_q}, where we consider the measured quark masses within $ 3\sigma $ evaluated at the $ M_Z $ scale.

Interestingly, upon inspecting the mass ratios of the first generation fermions in both the quark and lepton sector, it is found that the peak likelihood approximately satisfies $ \frac{m_{1}^f}{m_{3}^f}\sim 0.4\epsilon^f_{\textrm{max}} $. Unlike the second generation quarks, the mass distributions of the first generation quarks are very broad due to the number of variables within the anarchy perturbations. The down quark mass has been allowed freedom in the choice of its mass, as the down sector is the dominant source of the mixing in the CKM matrix in this scenario, which is  due to the its weakened hierarchy compared to the up sector. The central value of the mass ratio of the strange and bottom quarks has also been chosen at approximately $1 \sigma$ from the experimental central value to give improved consistency in the CKM mixing angle  $\theta_{23}$ prediction.

\begin{figure}
	\centering
	\begin{subfigure}
		\centering
		\topinset{\bfseries(a)}{\includegraphics[width=0.45\textwidth]{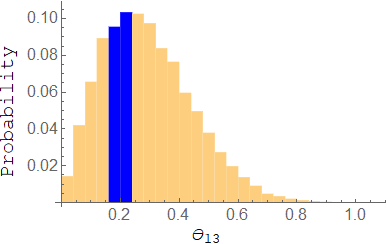}}{0.06in}{-.875in}
	\end{subfigure}
	\hfill
	\begin{subfigure}
		\centering
		\topinset{\bfseries(b)}{\includegraphics[width=0.45\textwidth]{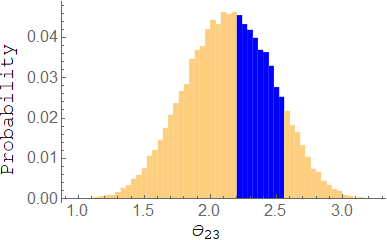}}{0.06in}{-.875in}
	\end{subfigure}
	\begin{subfigure}
		\centering
		\topinset{\bfseries(c)}{\includegraphics[width=0.45\textwidth]{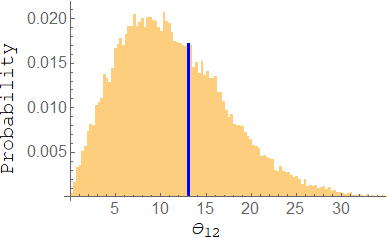}}{0.06in}{-.835in}
	\end{subfigure}
	\hfill
	\begin{subfigure}
		\centering
		\topinset{\bfseries(d)}{\includegraphics[width=0.45\textwidth]{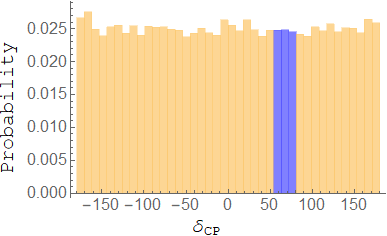}}{0.01in}{-.825in}
	\end{subfigure}
	\caption{Predicted mixing angles $\theta_{13}$, $\theta_{23}$, $\theta_{12}$, $\mathcal{CP}$ phase , respectively, in the quark sector, for the mass distributions given in Fig. \ref{mass_q}. The blue regions indicate the mixing angle ranges consistent with experiment within $3\sigma$ \cite{Tanabashi:2018oca}.}
	\label{angles_q}
\end{figure}

The stronger mass hierarchy present in the up sector relative to the down sector, means that the dominant source of the $ \theta_{12} $ and $ \theta_{13} $ mixing angles shall be induced by the properties of the down sector. The preferred values of the mixing angles $ \theta_{12} $ and $ \theta_{13} $ are dependent on the maximum size of the anarchy perturbations $\epsilon_\textrm{max}^d$. If it is assumed that the preferred down quark mass coincides with the experimental central value, it is found that the predicted $ \theta_{12} $ and $ \theta_{13} $ mixing angles are too small. If instead, the allowed magnitude of the random perturbations, $ \epsilon_d $, is increased, the observed mixing angles can be recovered, as depicted in Fig. \ref{angles_q}. In this scenario, where $ \epsilon^d=0.01 $, the down quark mass distribution preferences values larger than those within observational errors. Although this might seem like an issue, upon taking only mass values within this distribution that are consistent with observation, we find no change in the predicted angle distributions. Thus, the mixing angle predictions are not strongly dependent on what down quark masses are sourced from the mass distribution, rather they are dependent on the value of $\epsilon_\textrm{max}^d$. This indicates that selecting for the experimentally allowed range of the down quark mass is in effect selecting for the pattern of anarchy perturbations for a small contribution to $m_d$ while not applying constraints on other elements within the perturbation matrix that effect the mixing angles.  It has recently been suggested that QCD instanton effects may impact the masses of the light quark states, and as such there could be larger uncertainty in the mass of the up and down quarks than currently considered \cite{Bardeen:2018fej}. If so, this may reconcile the discrepancy between the preferred down quark mass with observation. In the case of the up quark, it's effects are not dominant in regards to the mixing angles so alterations to the allowed mass range should be easily accommodated. 

It should be noted that, if the down sector anarchy perturbations exhibit an  $\mathbb{S}_2 \times\mathbb{S}_2$ like structure, with the first 2 by 2 entries suppressed, it is possible to alleviate the tension of the down quark mass between experiment and prediction. Interestingly, the consistency of the CKM mixing angles with observation is simultaneously maintained. This serves to illustrate the patterns of down sector anarchy perturbations that lead to experimentally consistent results, even when considering an enlarged value of $\epsilon_\textrm{max}^d$.

\subsection{Neutrinoless Double Beta Decay}
\label{sec:nuDBD}

\begin{figure}[h]
		\centering
	\begin{subfigure}
		\centering
		\includegraphics[width=0.48\textwidth]{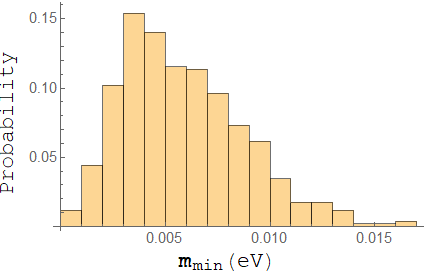}
	\end{subfigure}
				\hfill
	\begin{subfigure}
		\centering
		\includegraphics[width=0.48\textwidth]{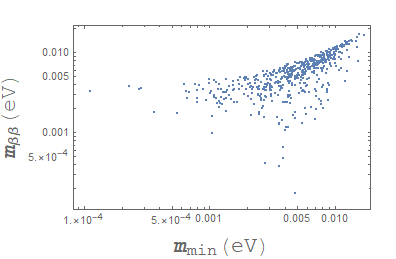}
		\end{subfigure}
\caption{Neutrinoless double beta decay predictions, where only points consistent with the observed neutrino mass difference ratio $ \frac{\Delta m_{21}^2}{\Delta m_{31}^2} $, $\theta_r$, and $\theta_s$ within $3 \sigma$ are selected. The mass of $m_3$ is fixed as $m_3\sim \sqrt{\Delta m_{31}^2}\sim 0.05$ eV, and all points are found to lie within the general case expectations. }\label{nuDBD}
\end{figure}

Neutrinoless double beta decay is a predicted experimental feature of models containing Majorana neutrinos which may be testable in the near future. Theories that provide unique predictions may be differentiated or ruled out by increased experimental precision of the measurement of this process. To determine the possibility for observability of this phenomena in our model we consider only those points in Fig. \ref{mass_l} that replicate the observed neutrino mass difference ratio $ \frac{\Delta m_{21}^2}{\Delta m_{31}^2} $, $\theta_r$, and $\theta_s$  within the $3\sigma$ constraints. The predicted effective Majorana mass $m_{\beta \beta}$ is depicted in  Fig. \ref{nuDBD}. Note, the predicted distribution is found to be mostly unchanged by additionally applying constraints on $\theta_r$.

Interestingly, we find that there is preference for values of $m_{\beta \beta}\sim 0.1-0.2 \cdot m_3$, which is within reach of future detectors when considering  $m_3\sim \sqrt{\Delta m_{31}^2}\sim 0.05$ eV. The potential observability of neutrinoless double beta decay in this model is enhanced due to the large values of $m_1$ and $m_2$ compared to $m_3$ that result from the maximum magnitude of the anarchy terms not being suppressed relative to the $\mathbb S_3$ symmetric terms which are disallowed by the $\mathbb Z_4$ symmetry.  Thus, measurement of neutrinoless double beta decay within this region could be a crucial test of this model.

\section{Conclusions}
\label{sec:conclusion}

The democratic flavor mixing pattern is an intriguing possibility for explaining the observed fermionic mass mixing structure in the standard model, due to the inherent mass hierarchy it exhibits and large mixing angles. Although it fails to correctly predict the mixing angles and light fermion masses, broken democracy may be a viable candidate. We have introduced both subleading $ \mathbb S^L_2 \times \mathbb S^R_2 $ symmetric perturbations and anarchy perturbations to the democratic mass matrix to possibly resolve these issues. The $ \mathbb S^L_2 \times \mathbb S^R_2 $ perturbations contribute predominantly to the second generation fermion masses, and are essentially fixed in size by Eq. (\ref{mass2}), while the anarchy perturbations are responsible for generating the light fermion masses. The analytical predictions exhibit interesting relations between the mixing angles and the second and third generation fermion mass ratios. Upon introducing anarchy perturbations and requiring the replication of the experimentally observed masses the PMNS mixing angles are found to be consistent with experiment, within $3\sigma$. This is made possible by the presence of the $ \mathbb{Z}_4 $ symmetry in the neutrino sector naturally leads to large anarchy perturbations relative to the quarks and charged leptons through forbidding the leading $ \mathbb{S}_{3} $ symmetric terms. In the quark sector, the predicted CKM mixing angles are not within the current $3\sigma$ bounds, when considering he central value of the measured quark masses. Upon giving freedom to the preferred down quark mass the observed mixing angles can be recovered. In this case, mixing angle predictions are independent of the choice of down quark mass and instead are dependent on $\epsilon^d_{\textrm{max}}$.  Considering the simple form of the mixing matrices and the minimal constraints assumed, namely the fermion masses, the consistency of the model predictions with the observed CKM and PMNS mixing angles is an intriguing result. The predictions for neutrinoless double beta decay also provide an interesting test of this model.

\section*{Acknowledgements}

This work is supported by World Premier International (WPI) Research Center Initiative, MEXT, Japan.
SFG is supported by JSPS KAKENHI Grant Number JP18K13536.
T. T. Y. is supported in part by the China Grant for Talent Scientific Start-Up Project and the JSPS Grant-in-Aid for Scientific Research No. 16H02176, and No. 17H02878 and by World Premier International Research Center Initiative (WPI Initiative), MEXT, Japan. T. T. Y. thanks to Hamamatsu Photonics.

\end{document}